\documentclass[aip,jcp,reprint,nofootinbib]{revtex4-1}
\usepackage{mathrsfs}
\usepackage{amsmath}
\usepackage{amssymb}
\usepackage{color}
\usepackage{graphicx}	% Include figure files
\usepackage{bm}% bold math
\usepackage{ulem} % texte barré
\newcommand{\be}{\begin{equation}}
\newcommand{\ee}{\end{equation}}
\newcommand{\bef}{\begin{figure}}
\newcommand{\eef}{\end{figure}}
\newcommand{\bea}{\begin{eqnarray}}
\newcommand{\eea}{\end{eqnarray}}
\newcommand{\brho}{\mbox{\boldmath${\rho}$}}

\DeclareMathOperator\erf{erf}
	     
\begin{document}
\title{DNA denaturation bubbles: free-energy landscape and nucleation/closure rates}
\author{Fran\c cois Sicard}
\thanks{Corresponding author: \texttt{francois.sicard@free.fr}.}
\author{Nicolas Destainville}
\author{Manoel Manghi}
\affiliation{Universit\'e de Toulouse, UPS, Laboratoire de Physique Th\'eorique (IRSAMC), F-31062 Toulouse, France, EU}
\affiliation{CNRS; LPT (IRSAMC); F-31062 Toulouse, France, EU}
%\date{\today}	     
%	     
\begin{abstract}
The issue of the nucleation and slow closure mechanisms of 
non superhelical stress-induced denaturation bubbles 
in DNA is tackled using coarse-grained MetaDynamics 
and Brownian simulations. A minimal mesoscopic model is used 
where the double helix is made of two interacting bead-spring 
rotating strands with a prescribed torsional modulus in the duplex state. 
We demonstrate that timescales for the nucleation (resp. closure) 
of an approximately 10~base-pair bubble, in agreement with experiments, 
are associated with the crossing of a free-energy barrier of $22~k_{\rm B}T$ 
(resp. $13~k_{\rm B}T$) at room temperature $T$. MetaDynamics allows us 
to reconstruct accurately the free-energy landscape, to show that 
the free-energy barriers come from the difference in torsional energy 
between the bubble and duplex states, and thus to highlight the limiting step, 
a collective twisting, that controls the nucleation/closure mechanism, 
and to access opening time scales on the millisecond range. 
Contrary to small breathing bubbles, these more than 4~base-pair bubbles 
are of biological relevance, for example when a preexisting state 
of denaturation is required by specific DNA-binding proteins.
\end{abstract}


\maketitle

\section{Introduction}
Although the DNA structure in double-helix is robust enough to enable 
the preservation of the genetic code, it is sufficiently loose to allow 
the formation of denaturation bubbles, \textit{i.e.} the cooperative opening 
of a sequence of consecutive base-pairs (bps), even at physiological temperature.
DNA opening is central in biological mechanisms such as replication, transcription, 
repair, or protein binding~\cite{Kornberg-Freeman1992,Phillips-Garland2009,Leger-PNAS1998, Kowalski-PNAS1988}. 
The magnitude of the interactions between two bases is of a few $k_{\rm B}T_0$ 
($T_0=300$~K is room temperature)~\cite{SantaLucia-PNAS1998, Poland-Academic1970}, 
and the base-pair opening is closely related to DNA elastic properties. 
Indeed, a denaturation bubble has much smaller bending and torsional moduli 
than the double helix one~\cite{Benham1979, Jeon-PRL2010, Manghi-JPCM2009,Dasanna-PRE2013}.
Considering the timescale of DNA replication and transcription 
(replication rates are roughly 1000~bp/s~\cite{Phillips-Garland2009}), 
the lifetime of large denaturation bubbles is expected to be on the order 
of $1~\mu$s to $1$ ms, as shown in \textit{in vitro} experiments~\cite{Altan-PRL2003}, 
where large bubble lifetimes of 20 to $100~\mu s$ have been observed, 
even for DNA constructs as small as 30-bps.

Various numerical and theoretical models have been proposed in the literature 
to account for the thermodynamic and dynamical properties of denaturation bubbles. 
DNA denaturation is tackled at different levels of coarse-graining and timescales, 
going from classical all-atom~\cite{Cheatham-JMB1996, Dixit-BJ2005, Kannan-PCCP2009} 
or coarse-grained (CG)~\cite{Sayar-PRE2010, Ouldridge-JCP2011, Savelyev-PNAS2010, Englander-PNAS1980, Zeida-PRE2012} 
molecular dynamics simulations, to mesoscopic models focusing either on 
the inter-strand distance dynamics~\cite{Peyrard-PRL1989, Barbi-PL1999, Jeon-JCP2006}, 
or on the bubble size dynamics using the Poland-Scheraga model~\cite{Bar-PRL2007, Fogedby-PRL2007}. 
However, these approaches did not consider explicitly the twist dynamics and/or were not able 
to reach the 100~$\mu$s timescale for long enough sequences.
Nonetheless, Mielke et al.~\cite{Mielke-JCP2005} studied the interplay between 
denaturation and writhe, but the applicability of this model was limited to 
non-equilibrium conditions imposed by the dynamic introduction of torsional stress.

\begin{figure}[b]
 \includegraphics[width=0.9 \columnwidth]{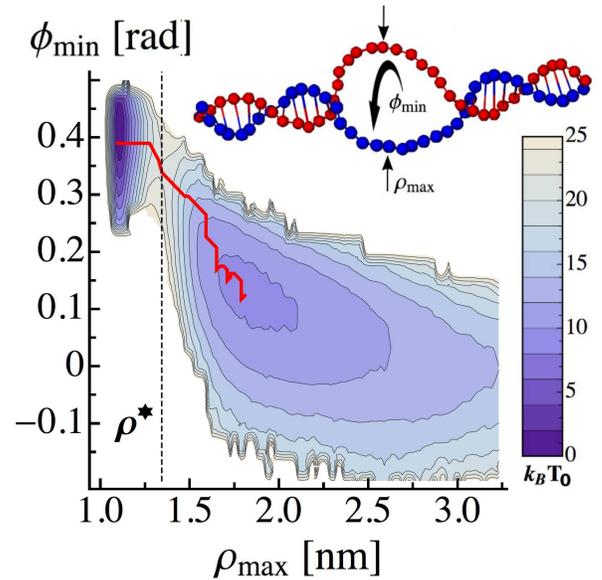}
 \caption{Free-energy surface associated with the bubble closure/nucleation mechanism projected along two observables ($\beta_0\kappa_\phi = 580$): the maximal distance between paired bases $\rho_{\max}$ and the minimal twist angle between successive bps, $\phi_{\min}$  (see inset). The saddle point is located at $\rho^*=1.35$ nm. The typical minimal free-energy path is shown in red color, and the 
 contour lines are every $2k_BT_0$.}
 \label{fig1}
\end{figure}
\begin{figure*}[t]
\includegraphics[width=0.85 \textwidth]{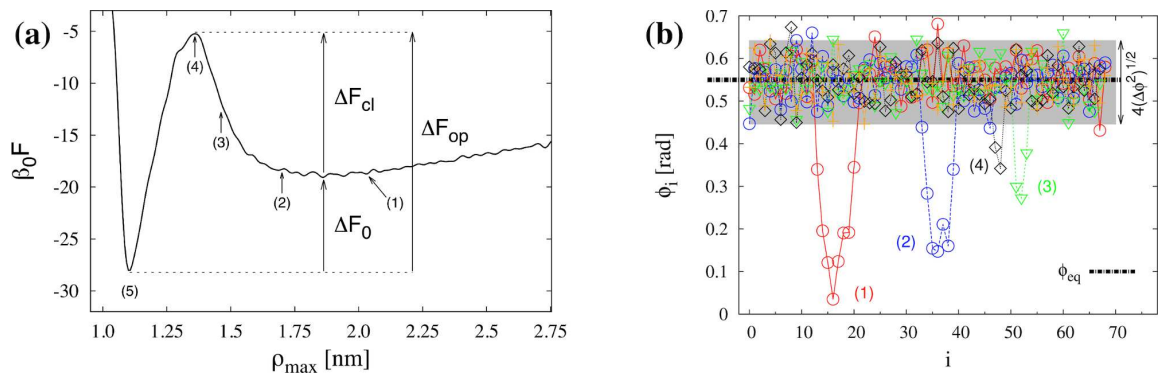}
 \caption{(a)~Free-energy profile associated with the opening/closure mechanism for $\beta_0\kappa_\phi = 580$. 
(b)~Evolution of the twist angle profile $\phi_i(t)$ for different bubble configurations labelled from~(1) to~(4) in~(a).  The fluctuations around the equilibrium value of 0.55~rad [configuration~(5)] are represented in grey.}
 \label{fig2}
\end{figure*}
In an attempt to understand the bubble dynamics at the microsecond scale 
in the absence of supercoiling, a simple CG model, 
introduced in Ref.~\onlinecite{Dasanna-PRE2013}, showed that long closure times 
are related to the crossing of a free-energy barrier from a metastable bubble 
of $\approx 10$~bps to the duplex state. This result accounts for 
the Arrhenius laws measured experimentally in Ref.~\onlinecite{Altan-PRL2003}. 
Furthermore the role of torsion was highlighted. However, despite 
the relative simplicity of the model, many processes are still slow 
to equilibrate due to the presence of a large free-energy barrier, 
which made the closure a rare event and opening rates inaccessible.

In the present work, Well-Tempered MetaDynamics (WT-metaD)~\cite{Laio-PNAS2002, Laio-RPP2008, Barducci-PRL2008, Dama-PRL2014}, \textit{i.e.} biased numerical simulations, allow us to study accurately, 
without these limitations, the thermodynamic and dynamical properties 
of denaturation bubbles by constructing the whole free-energy landscape, 
shown in Fig.~\ref{fig1}. 
Whereas most of the theoretical works focused on the dynamics of small breathing 
bubbles~\cite{Peyrard-PRL1989,Barbi-PL1999,Anil-Thesis2003,Warmlander-Biochem1999}, 
we show that the torsion and bending of strands play a pivotal role in the nucleation 
and closure of bubbles of sizes larger than 4 bps, 
and which are not superhelical stress-induced~\cite{Mielke-JCP2005, Bauer-JMB1993}. 
These are of course the denaturation bubbles which are of biological relevance 
when a long-lived preexisting state of denaturation 
may be required by specific DNA-binding proteins~\cite{Vlaminck-MC2012}.
We highlight that the associated free-energy barrier is of elastic nature, 
related to single-strand bending and boundary conditions at the bubble extremities, 
which induce a  torsional modulus, $\kappa_\phi^*(L)$, of the whole bubble of length $L$. 
This non-vanishing  $\kappa_\phi^*(L)$ is shown to be at the origin of 
the free-energy barriers. For physically relevant model parameters, 
we find an opening free-energy barrier $\Delta F_{\rm op}=22~k_{\rm B}T_0$ 
and a closure one $\Delta F_{\rm cl}=13~k_{\rm B}T_0$. The respective mean 
opening and closure times are measured numerically~\cite{Tiwary-PRL2013, Salvalaglio-JCTC2014}: 
$\tau_{\rm op}=15\pm3$~ms and $\tau_{\rm cl}=40\pm9~\mu$s. We emphasize that 
the so-obtained thermodynamic and dynamical properties are in agreement 
with experiments~\cite{Altan-PRL2003} and biological 
mechanisms~\cite{Qi-eLife2013, Myong-Science2007}. 

\section{Numerical model}
We use the DNA model of Ref.~\onlinecite{Dasanna-PRE2013}, where the two single strands 
are modeled as freely rotating chains (FRC)~\cite{Grosberg-AIP1994} 
of $N = 70$ beads of diameter $a =0.34$~nm 
with a AT-rich region of $50$ bps clamped by GC regions 
of $10$ bps~\cite{footnote}.  
These beads interact through two terms: a Morse potential mimicking 
the inter-strand hydrogen-bonding and an \textit{effective} intra-strand stacking interaction 
between the base-pairs modeled through a bare torsional modulus, 
$\kappa_{\phi,i}(\rho_i)$, that depends on the distance between 
complementary bases, $\rho_i = |\brho_i| = |\textbf{r}_i^{(1)}-\textbf{r}_i^{(2)}|$ 
with $\textbf{r}_i^{(j)}$ the position of bead $i$ on strand $j$, 
and vanishes for fully separated strands. 
The evolution is governed by the overdamped Langevin equation. 
The full Hamiltonian and the details of the numerical implementation 
and of the parameter values are given in 
the appendix. Note that, as compared 
to Ref.~\onlinecite{Dasanna-PRE2013}, model parameters are modified to account for 
realistic opening times, which were not previously accessible 
without WT-metaD simulations, but without any direct 
\textit{a priori} on the closure times. The value of the \textit{bare} 
torsional modulus, $\kappa_\phi$, in the duplex state, is chosen 
so that its actual torsional modulus, $\kappa_{\phi, \textrm{ds}}^*$, 
is close to $450~k_{\rm B}T_0$, consistent with experimental 
values~\cite{Bryant-Nature2003}. The equilibrium properties of this model 
are described in Ref.~\onlinecite{Dasanna-PRE2013} and in the appendix. This model showed that 
the twist dynamics plays a key role in the closure of pre-equilibrated 
large bubbles, which occurs in two steps~\cite{Dasanna-PRE2013}: 
First, the large flexible bubble quickly winds from both ends 
(\textit{zipping} regime~\cite{Dasanna-EPL2012, Dasanna-PRE2013}), 
thus storing bending and torsional energy in the bubble, which stops 
when it reaches a size of $\approx 10$~bps (see Fig.~\ref{fig1}). 
For large $\kappa_\phi$ and $N$, or clamped ends, the ultimate closure 
of this metastable bubble is then \textit{temperature-activated}~\cite{Dasanna-PRE2013}.

WT-metaD enhances the sampling of the conformational space of a system 
along a few selected degrees of freedom, named \textit{collective variables} (CVs), 
and reconstructs the equilibrium probability distribution, 
and thus the free-energy landscape, as a function of these CVs (see Fig.~\ref{fig1}).
The chosen CVs must mainly account for the relevant barriers 
associated with CG variables on which the free-energy dependence 
is the most important. Several observables come out naturally to describe the metastable state 
and the transition to the closed state: (1)~the length $L(t)$ of the bubble, 
\textit{i.e.} the number of opened base-pairs, (2)~the width $\rho_{\max}(t)$ 
of the bubble, \textit{i.e.} the maximal distance between paired bases, 
(3)~the average twist angle per bp in the bubble~\cite{Dasanna-PRE2013}, 
$\Delta \phi(t)=\langle\phi_i(t)\rangle_{i\in \textrm{bubble}}$, 
where the local twist $\phi_i \equiv \arccos\Big( \frac{\brho_i.\brho_{i+1}}{\rho_i \rho_{i+1}} \Big)$ 
is the angle between two consecutive 
base-pair vectors, and (4)~the minimal twist angle inside the bubble, 
$\phi_{\min}(t)=\min_{i\in \textrm{bubble}} \phi_i(t)$. 
For numerical efficiency, we choose the width $\rho_{\max}$ as CV 
to bias the dynamics. Computational details are given in the appendix. 
To explore the twist dynamics, we choose to follow the evolution 
of $\phi_{\min}(t)$ instead of $\Delta \phi(t)$, as the latter 
is very noisy for small bubble sizes and is not defined at all 
in the closed state.

\section{Results}
In Fig.~\ref{fig2}(a) is shown the free-energy profile associated 
to the closure mechanism for $\beta_0\kappa_\phi = 580$ ($\beta_0^{-1}=k_{\rm B}T_0$) 
along the width $\rho_{\max}(t)$ of the bubble. 
A \textit{closure} free-energy barrier, $\Delta F_{\rm cl}$, 
of approximately $14~k_{\rm B}T_0$ separates the metastable basin 
associated with the denaturation bubble ($\rho_{\max}\geq 1.35$ nm) 
from the closed state basin ($\rho_{\max}\approx 1.1$ nm). 
These two basins are well separated by a standard free-energy of formation 
$\Delta F_0 \approx 8~k_{\rm B}T_0$, defining the \textit{opening} free-energy barrier, 
$\Delta F_{\rm op}\equiv \Delta F_0 + \Delta F_{\rm cl} \approx 22~k_{\rm B}T_0$, 
associated with the nucleation mechanism. The corresponding evolution 
of the twist angle profile $\phi_i(t)$ [and thus the minimal twist $\phi_{\min}(t)$] 
in the bubble is shown in Fig.~\ref{fig2}(b). The minimal twist 
inside the bubble increases when the bubble closes, going from an average value 
of 0.1~rad (configuration 1) to the ds one, 0.45~rad (configuration 5). 
In addition to the bubble diffusion along the dsDNA axis, we clearly see 
that the evolution of $\phi_{\min}(t)$ confirms a \textit{collective twisting} 
mechanism associated with the existence of the free-energy barrier, 
\textit{i.e.} $\phi_{\min}(t)$ decreases as $L(t)$ decreases. This mechanism is drastically 
different from the one at play during the zipping process for which the system 
is controlled by a \textit{processive twisting}, {\textit{i.e.} $L(t)$ decreases while keeping 
$\phi_{\min}(t) \approx 0$ at the center of the bubble~\cite{Dasanna-PRE2013}. 
Let us note that switching from AT- to GC-rich region in the model 
does not change qualitatively the physics of nucleation/closure mechanism, 
mainly affecting $\Delta F_{\rm op}$.

To go further, we show in Fig.~\ref{fig1} the free-energy surface 
projected along two observables ($\rho_{\max}$, $\phi_{\min}$),  
and reconstructed using the \textit{reweighing technique} of 
Bonomi \textit{et al.}~\cite{Bonomi-JCC2009}. A typical minimal free-energy path 
is also shown (in red in Fig.~\ref{fig1}) and displays two different regimes. 
Starting from the metastable basin ($\rho_{\max} \approx 2$ nm), 
the system is driven by a \textit{collective twisting} (the oblique part of the red path in Fig.~\ref{fig1}) 
up to the saddle point $\rho^*$. The end of the evolution, ($\rho_{\max} < \rho^*$) 
shows a plateau at $\phi_{\min} = \phi_{\min}^{\rm eq} \approx 0.4$. 
This is characteristic of a breathing bubble, \textit{i.e.} the fast opening 
and closure of a few bps on nanoseconds without modification of 
the conformation of the whole chain. It precises the previous notion 
of transient (or breathing) bubble~\cite{Peyrard-PRL1989, Barbi-PL1999,Anil-Thesis2003,Warmlander-Biochem1999}, 
and corresponds to bubbles of size $L(t)\leq 4$~bps.
To ensure the reliability of the model with experiments, 
we study in Fig.~\ref{fig3} the dependence of the closure free-energy barrier, 
$\Delta F_{\rm cl}$, and the free-energy of formation, $\Delta F_0$, 
on $\beta_0\kappa_\phi$. As anticipated, the free-energy barriers 
$\Delta F_{\rm cl}$ (resp. $\Delta F_0$), increases (resp. decreases) 
for increasing values of $\beta_0\kappa_\phi$, scaling affinely 
in an energy range in agreement with experimental 
observations~\cite{Altan-PRL2003} and biological 
mechanisms~\cite{Qi-eLife2013, Myong-Science2007}. Therefore, the opening 
free-energy barrier, $\Delta F_{\rm op}$, increases more slowly than $\Delta F_{\rm cl}$.
\begin{figure}[t]
 \includegraphics[width=1.0 \columnwidth]{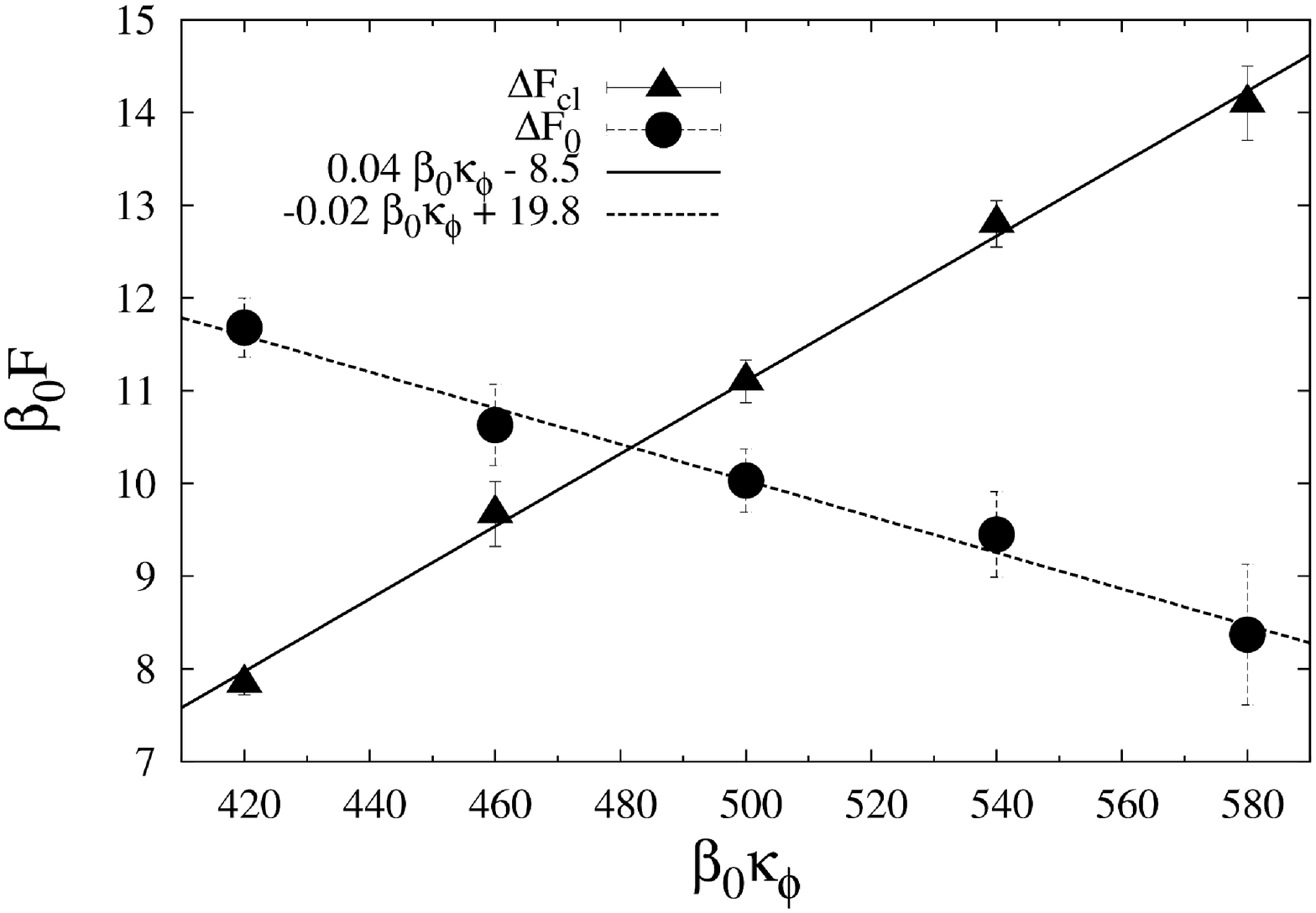}
 \caption{Evolution of the closure free-energy barrier, $\Delta F_{\rm cl}$ (triangles), 
 and the free-energy of formation, $\Delta F_0$ (circles), for increasing values 
 of the \textit{bare} torsional modulus, $\kappa_\phi$. $95\%$ confidence intervals are also 
 provided.}
 \label{fig3}
\end{figure}

Let us now explain the origin of these free-energy barriers. 
Although the mesoscopic model imposes by hand a vanishing torsional modulus 
in the ssDNA state, the free-energy barrier is actually related 
to geometrical constraints: the 2 single strands in the bubble are strongly connected 
to the double-stranded domain at the edges of the bubble.
Indeed, strand stretching and inter-strand interaction lead to an energetic cost associated with the bubble twist. 
The system can thus be seen as 2 rigid dsDNA arms connected by an effective joint of torsional 
rigidity $\kappa_{\phi}^*(L)$ (see Fig.~\ref{figA7} in the appendix).
This non-vanishing $\kappa_{\phi}^*$ is responsible for the stop of the zipping process.
Actually, the \textit{collective twisting} mechanism associated with 
the bubble closure is central for the effective joint representation: 
the double-stranded domains at the edges of the bubble 
are free to rotate around their own axis to relax the torsional constraint, 
but not free to rotate relative to one another (as it is the case in the zipping process).
This torsional modulus is measured, considering the equipartition theorem, 
as $\kappa_{\phi}^* = k_{\rm B} T/ \langle (\Phi - \langle\Phi\rangle)^2 \rangle$, 
where $\Phi \equiv \sum_{i\in \textrm{bubble}} \phi_i$ is the twist angle 
measured \textit{consecutively} between the bps defining each extremity 
of the bubble. It is characterized by a non-trivial power law behaviour, 
$\kappa_{\phi}^*(L) \propto L^{-\alpha}$ with $\alpha = 2.2 \pm 0.1$, 
valid down to $L \approx 3$~bps that corresponds to the breathing bubble regime. 

The origin of the free-energy barrier is indeed related to the finite value 
of $\kappa_{\phi}^*(L)$ in the metastable bubble and the crossover between 
two minima  for the minimal twist angle, $\phi^{\rm eq}_{\min}$ and 0. 
Using a mean-field approximation where we consider only the bp located 
at the center of the bubble, and noting $\rho$ the distance between 
the two pairing bases and $\phi$ its twist, we write the following energy:
\be
%\mathcal{H}(\rho,\phi)=V_{\rm Morse}(\rho)+ \frac{\kappa_\phi(\rho)}2(\phi-\phi_{\min}^{\rm eq})^2+ \frac{\kappa_{\phi}^*(\rho)}2\phi^2~,
\mathcal{H}(\rho,\phi)=V_{\rm Morse}(\rho)+ \left\{ 
\begin{array}{l l}
\frac{\kappa_\phi(\rho)}2(\phi-\phi_{\min}^{\rm eq})^2 & \, \text{for $\rho\leq\rho_b$}\\
\frac{\kappa_{\phi}^*(\rho)}2\phi^2~& \, \text{for $\rho>\rho_b$}\\ 
\end{array} \right.
\ee
with $\rho_b\simeq 1.2$ nm and where $V_{\rm Morse}$ is the Morse potential, and the torsional 
 energy has a bending modulus which depends on the base-pair state: 
$\beta_0 \kappa_\phi\simeq580$  
and $\kappa_{\phi}^*(\rho)\simeq L(\rho)^{-\alpha}$ 
(the torsional potential is smoothed with error functions near $\rho_b$).  
Note that the dependence of $L$ on $\rho$ is almost linear 
in the metastable state (see Fig.~\ref{figA6}(a) in the appendix). 
The free-energy surface $\mathcal{H}(\rho,\phi)$ is projected along 
the two observables ($\rho$, $\phi$) in Fig.~\ref{fig4}. We observe 
a landscape very similar  to the one shown in Fig.~\ref{fig1}. 
Moreover, comparing Fig.~\ref{fig4}(a) and Fig.~\ref{fig4}(b) 
clearly highlights the role of $\kappa_{\phi}^*$ in the occurrence 
of the metastable state and the saddle point. Of course this simple model 
does not account explicitly for the cooperativity between bubble bps 
and thus yields a crude estimate of free-energy values. Nevertheless, 
it illuminates the role played by the torsional energy in the closure mechanism 
in the absence of superhelical stress-induced constraint.

\begin{figure}[t]
\includegraphics[width=1.0 \columnwidth]{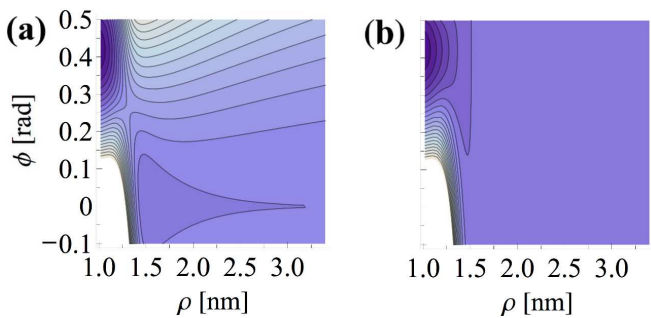}
 \caption{ 
 Energy surfaces of $\mathcal{H}(\rho,\phi)$ associated with the closure mechanism 
 (see text) for $\beta_0\kappa_\phi = 580$, and (a)~$\kappa_{\phi}^* \neq 0$ 
 (to be compared with Fig.~\ref{fig1}) and (b)~$\kappa_{\phi}^* = 0$, for which 
 the metastable basin is replaced with a flat free-energy landscape. 
 We use the same free-energy scale as in Fig.~\ref{fig1}.}
 \label{fig4}
\end{figure}
\begin{figure}[b]
 \includegraphics[width=1.0 \columnwidth]{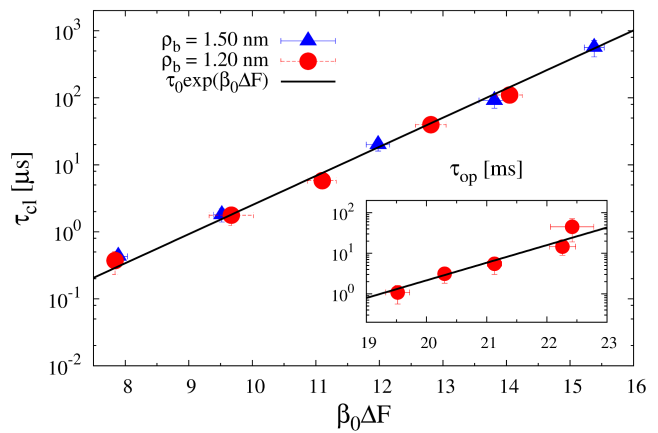}
 \caption{Mean transition times, $\tau_{\rm cl}$ and $\tau_{\rm op}$ (closure and nucleation), 
 at room temperature $T_0$, as a function of the height of the free-energy barrier 
 $\Delta F_{\rm cl}$, for $\rho_b = 1.20$ nm (circles) and $\rho_b = 1.50$ nm (triangles).}
 \label{fig5}
\end{figure}
Considering the recent method of Tiwary and Parrinello~\cite{Tiwary-PRL2013, Salvalaglio-JCTC2014}, 
we extend the standard application scope  of MetaDynamics in order to 
estimate the rate of transition between the metastable and the closed states 
(see computational details in the appendix). In Fig.~\ref{fig5} is shown 
the evolution of the mean transition times $\tau_{\rm cl}$ and $\tau_{\rm op}$ (inset) 
associated with the closure and  nucleation mechanisms, as a function of the barrier heights.
As expected from the thermodynamic analysis above, $\tau_{\rm cl}$ and $\tau_{\rm op}$ 
increase with the height of the free-energy barriers, \textit{i.e.} the value 
of the torsional  modulus $\beta_0 \kappa_\phi$. Furthermore, Fig.~\ref{fig5} 
displays an \textit{Arrhenius-like} exponential dependence of the mean transition times 
$\tau_{\rm cl}$ and $\tau_{\rm op}$ on the free-energy of activation, 
$\tau = \tau_0 \exp(\Delta F/k_{\rm B}T)$, which follows from Kramers 
theory~\cite{Hanggi-RMP1990}. This exponential dependence remains 
unchanged with varying values of $\rho_b$. Actually, in the strong friction regime 
of interest here, one has $\tau_0 = 2\pi\zeta/(\omega_{\rm met}\omega_{\rm TS})$, 
where $\omega_{\rm met}$ (resp. $\omega_{\rm TS}$) is the angular frequency 
inside the metastable basin (resp. the transition state), and $\zeta$ the friction 
coefficient. The angular frequency $\omega_{\rm TS}$ (and consequently $\tau_0$) 
depends on the choice of the parameters controlling the balance between the hydrogen-bonding 
and effective stacking interactions (see the appendix). 
Considering Fig.~\ref{figA4}(b) in the appendix, we see that a change of the values 
of $\rho_b$ does not affect significantly the shape of the transition state 
and the metastable basin, \textit{i.e.} the angular frequencies 
$\omega_{\rm TS}$ and $\omega_{\rm met}$, only modifying the height 
of the free-energy barriers. Consequently, this allows us to compare 
our dynamical analysis,with the experimental result of Altan-Bonnet 
\textit{et al.}~\cite{Altan-PRL2003},  associated with a characteristic time scale 
$\tau_{\rm{met}}^{\rm exp}\simeq 50~\mu$s. One then deduces from Fig.~\ref{fig5} 
and Fig.~\ref{fig3}, $\beta_0\kappa_\phi = 540$, which is significantly 
different from the value of Ref.~\onlinecite{Dasanna-PRE2013}, because WT-metaD helped us 
to better set the model parameters, and consistent with hydrogen 
exchange measurements~\cite{Englander-PNAS1980}. This value corresponds to 
an activation free-energy $\Delta F_{\rm op} \approx 22~k_BT_0$ 
and a characteristic time scale $\tau_{\rm op} \approx 15$ ms, for nucleation. 
Finally, these opening and closure timescales lead to an equilibrium constant 
$K(T_0) \equiv \tau_{\rm cl}(T_0)/\tau_{\rm op}(T_0) \approx 0.003$, comparable 
with experiments~\cite{Bonnet-PNAS1998, Altan-PRL2003, Englander-PNAS1980}. 
Furthermore, to assert the good matching of the thermodynamic properties of 
our model with experiments, we simulated the hairpin of Ref.~\onlinecite{Bonnet-PNAS1998} 
and found similar melting temperatures (see the appendix). 
Playing on the model parameters will enable us in a mean future to adjust the thermodynamic 
and dynamical quantities to different experimental contexts 
(\textit{e.g.} sequences, ionic strength).

\section{Discussion and Perspectives}
To go further, it would be interesting to take explicitly into account 
the role of base stacking in the single-stranded domain. 
Actually, it is another notable contribution to the energy of stabilization of ssDNA. 
Therefore, the classical approach considering the ssDNA as a FRC and isotropic hydrogen bonding 
remains questionable~\cite{Sponer-Biopolymers2002}. However, improving our understanding of particular phenomenon 
with mesoscopic models necessarily reduces the question to fundamental aspects. 
The use of all-atom simulations, or CG models considering explicitly stacking interaction 
in the single-stranded domain~\cite{Ouldridge-JCP2011, Zeida-PRE2012}, coupled to metadynamics, 
would be interesting, as a second step, to support our mechanism.
Furthermore, the role of hydrodynamics interactions~\cite{Manghi-SM2006}, 
which might accelerate the closure, would be also relevant.
Finally, let us comment on the potential biological implications 
of this work. While strand separation plays a pivotal role 
in many biological processes, such as replication and transcription, 
it is commonly accepted that these phenomena are driven primarily by the stresses 
that are imposed by DNA supercoiling through polymerase 
actions~\cite{Liu-PNAS1987, Michelotti-MCB1996, Kouzine-Cell2013}. 
However, some biological mechanisms, such as homology recognition~\cite{Vlaminck-MC2012}  
or cruciform extrusion~\cite{Bikard-MMBR2010}, are believed to be dependent of spontaneous DNA-breathing 
dynamics~\cite{Alexandrov-PLOS2009}. Therefore, even if 
negative supercoiling of the dsDNA is assumed to strongly promote the frequency 
of occurrence and lifetime of DNA-breathing bubbles~\cite{Jeon-PRL2010,Jeon-BJ2008}, 
there is no definitive evidence that the mechanism highlighted in the present work cannot 
occur in the absence of supercoiling. It might also be possible that this mechanism 
takes place to get behind biological mechanisms where DNA would not undergo sufficient 
torsional stress for bubble nucleation. 
This roadmap will be considered in the near future.

%\smallskip
\section*{Acknowledgments}
We acknowledge A.K.~Dasanna and P.~Rousseau for useful discussions. F.S. thanks J.~Cuny and P.~Tiwary for fruitful discussions concerning the PLUMED plugin~\cite{Bonomi-CPC2009} and for several very useful suggestions. We acknowledge financial support from the Agence Nationale de le Recherche (Grant No. ANR-11-NANO-010-01).

\section*{Appendix}
\appendix
%\counterwithin{figure}{section}

\section{Numerical Model}

As discussed in Ref.~\onlinecite{Manghi-JPCM2009}, our mesoscopic DNA model consists 
in two interacting bead-spring chains each made of 
$N = 70$ beads (of diameter $a = 0.34$ nm) at position $\textbf{r}_i$, with a AT-rich region of $50$ bps in the middle, 
and a GC region of $10$ bps at each extremity. The Hamiltonian is 
$\mathcal{H} =\mathcal{H}_{el}^{(1)} + \mathcal{H}_{el}^{(2)} + \mathcal{H}_{tor} + \mathcal{H}_{int}$, 
where the first two contributions are elastic energies of the strands $j=1,2$ which include both 
stretching and bending energies
\begin{equation}
\mathcal{H}_{el}^{(j)} = \sum_{i=0}^{N-1} \frac{\kappa_s}{2}(r_{i,i+1}-a_\textrm{ref})^2 
+ \sum_{i=0}^{N-1}\frac{\kappa_\theta}{2}(\theta_i-\theta_\textrm{ref})^2.
\end{equation}
The stretching modulus, $a^2\beta_0 \kappa_s = 100$, is a compromise between numerical efficiency 
and experimental values~\cite{Hugel-PRL2005}, where $\beta_0^{-1} = k_B T_0$ is the thermal energy,
$T_0 = 300$ K is the room temperature, and $a_\textrm{ref}=0.357$ nm. The bending modulus is large, $\beta_0 \kappa_\theta = 600$, 
to maintain the angle between two consecutive tangent vectors along each strand $\theta_i$ to the 
fixed value $\theta_\textrm{ref} = 0.41$ rad (see Fig.~\ref{figA1}). Each strand is thus modeled as a freely rotating chain (FRC). 
The third and fourth terms of $\mathcal{H}$ are the torsional energy and hydrogen-bonding interactions, respectively. 
The torsional energy is modeled by a harmonic potential
\begin{equation}
\mathcal{H}_{tor} = \sum_{i=0}^{N-1} \frac{\kappa_{\phi,i}}{2}(\phi_i-\phi_\textrm{ref})^2 ,
\end{equation}
where $\phi_i$ is defined as the angle between two consecutive base-pair vectors
$\brho_i \equiv \textbf{r}_i^{(1)}-\textbf{r}_i^{(2)}$ 
and  $\brho_{i+1}$ ($\phi_\textrm{ref} = 0.62$ rad). 

The stacking interaction between base pairs is modeled through a $\kappa_{\phi,i}$ that depends on 
the value of the \textit{bare} dsDNA torsional modulus $\kappa_\phi$, and the distances between complementary bases,
$\kappa_{\phi,i} = \kappa_\phi [1-f(\rho_i)f(\rho_{i+1})]$, where 
\begin{equation} 
f(\rho_i) = \frac{1}{2}\Big[1+\erf\Big(\frac{\rho_i -\rho_b}{\lambda'}\Big)\Big], 
\label{stacking}
\end{equation}
and $\rho_i =|\brho_i|$. Hence, $\kappa_{\phi,i} = \kappa_\phi$ in the dsDNA state and $\kappa_{\phi,i} = 0$ 
in the ssDNA one. The actual values in the dsDNA state after equilibration, $\kappa^*_{\phi,\rm ds}$, 
are however different from the prescribed values, $\kappa_{\phi}$, due to thermal fluctuations and non-linear potentials 
entering the Hamiltonian. 
Nevertheless, Fig.~\ref{figA2} underlines the linear correlation between the prescribed and actual values of the 
torsional modulus, which is representative of the robustness of the mesoscopic model.
To compare the mesoscopic model with experiments, we study the dependence on the value of the  
torsional modulus $\kappa_{\phi,i}$ of the free-energy barrier. In Fig.~\ref{figA3} is shown the evolution 
of the one-dimensional free-energy profile along the width $\rho_{\max}$ for various \textit{bare} torsional modulus, $\kappa_\phi$. 
As we could expect from a preliminary study \cite{Dasanna-PRE2013}, the height of the barrier increases 
as a function of $\beta_0\kappa_\phi$. This increase of the height of the free-energy barrier remains local 
(around the saddle region), and does not affect significantly the shape of the metastable basin.
\begin{figure*}
 \includegraphics[width=0.85 \textwidth, angle=-0]{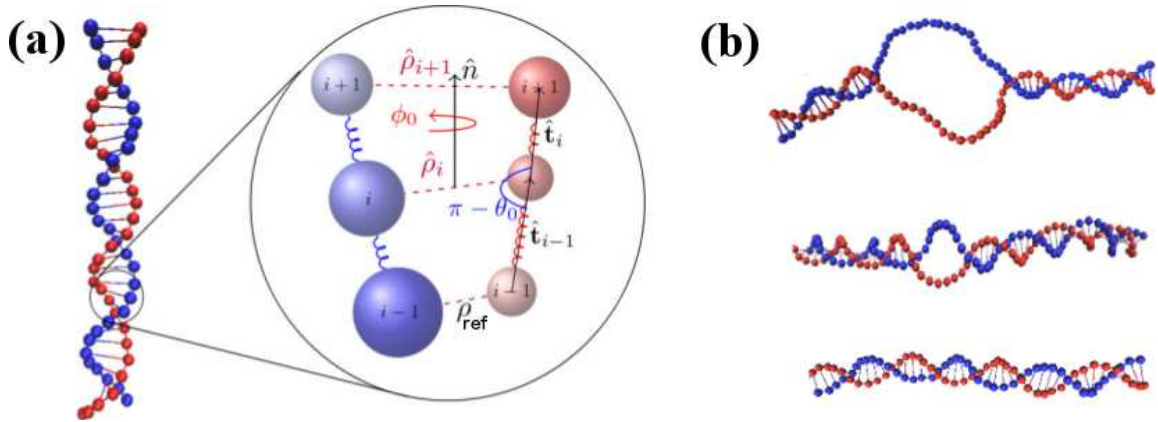}
 \caption{(a) Snapshot of an equilibrated double helix (from Ref.~\onlinecite{Dasanna-PRE2013}). 
 The bending angle along each strand is $\theta_\textrm{ref}$, $\rho_\textrm{ref}$ is the equilibrium base-pair distance 
 and $\hat{n}$ is the helical axis around which twist is defined. 
 The imposed equilibrium twist between successive pairs is $\phi_\textrm{ref}$. 
 (b) Graphic representation of the evolution of the bubble closure/nucleation.}
 \label{figA1}
\end{figure*}

As shown in Fig.~\ref{figA4}, the thermodynamic properties of the model are also sensitive to the values of the parameters 
$\lambda'$ and $\rho_b$ defined in Eq.~(\ref{stacking}); the latter playing a crucial role on the height 
of the energy barrier on both sides of the saddle region. 
Let us note, however, that a change in these values does not change qualitatively the \textit{physics} 
of the model, \textit{i.e.} the mechanism of nucleation and closure of long denaturation bubble. 
Considering preliminary simulations, we  have chosen $\lambda'=0.15$ nm 
and a range of $\rho_b \in [1.20~\rm{nm}, 1.50~\rm{nm}]$. In this work, we mainly focus on $\rho_b=1.20$ nm, 
which yields thermodynamic and dynamical properties in good agreement with biophysical mechanisms, 
\textit{i.e.} to account for realistic opening times, but without any direct \textit{a priori} on the closure times 
that emerge from Libchaber's experiment~\cite{Altan-PRL2003}. 
Playing on the values of $\rho_b$ and $\kappa_\phi$ would enable one to adjust the values of $\Delta F_0$ and $\Delta F_{\rm cl}$ to different experimental contexts (\textit{e.g.}, sequences, ionic strength).
\begin{figure*}
\includegraphics[width=0.85 \textwidth,]{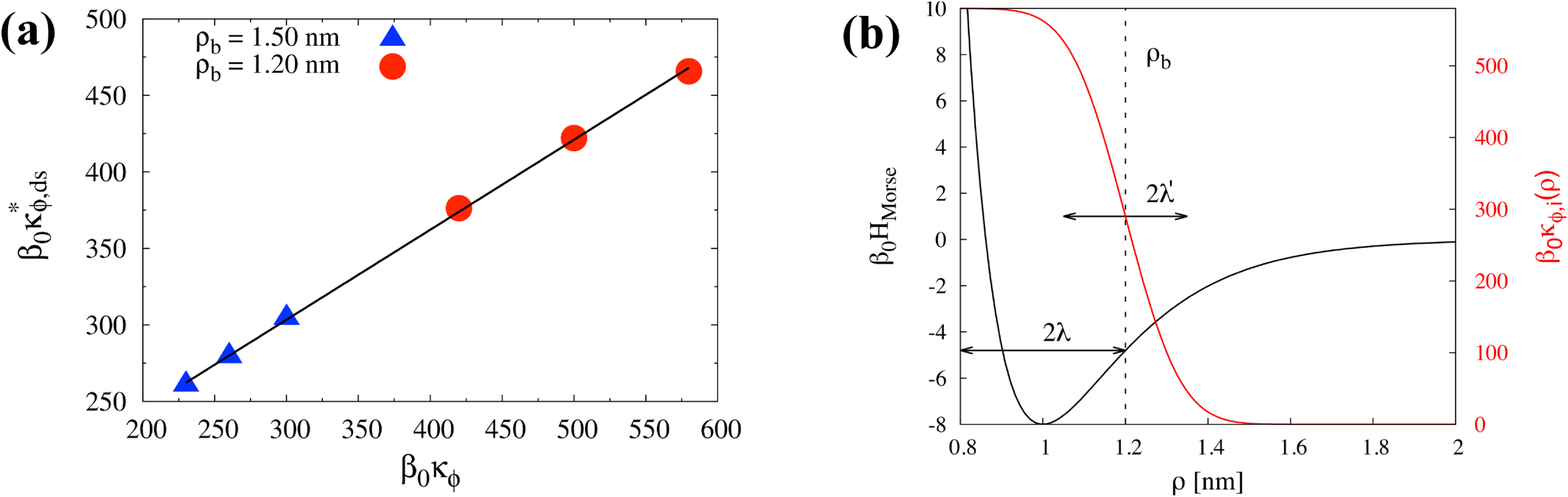}
 \caption{(a) Evolution of the actual torsional modulus, $\kappa^*_{\phi,\rm ds}$, of dsDNA, as a function 
 of the \textit{bare} torsional modulus, $\kappa_\phi$, for $\rho_b = 1.50$ nm (triangles) and 
 $\rho_b = 1.20$ nm (circles): $\kappa^*_{\phi,\rm ds} = (0.59 \pm 0.01) \kappa_\phi + (127 \pm 3)$. 
 As expected from the definition of the stacking interaction in Eq.~(\ref{stacking}), 
 the value of $\rho_b$ does not play a role in the duplex state. (b) Representation of the 
 hydrogen-bonding interaction modeled by a Morse potential (black curve) and the stacking interaction between base pairs (red curve). 
 The choice of the internal parmeters $(\lambda, \rho_\textrm{ref})$ and $(\lambda', \rho_b)$ fixes the 
 transition between stacked and unstacked regime taking into account local denaturation.}
 \label{figA2}
\end{figure*}
The hydrogen-bonding interaction is modeled by a Morse potential 
\begin{equation}
\mathcal{H}_{int} = \sum_{i=0}^{N-1} A (e^{-2\frac{\rho_i-\rho_\textrm{ref}}{\lambda}} -2e^{-\frac{\rho_i-\rho_\textrm{ref}}{\lambda}}) ,
\end{equation}
where $\rho_\textrm{ref}=1$ nm, $\lambda=0.2$ nm, and $\beta_0 A=8$ and $12$ for AT and GC bonding, respectively, 
as in Refs.~\onlinecite{Dasanna-EPL2012, Dasanna-PRE2013}.
The fitted values for the dsDNA persistence length and the pitch are $\ell_{\rm ds}\simeq160$~bps and $p = 12$~bps for the relevant range of $\beta_0\kappa_\phi$ we are interested in, which are comparable to the actual dsDNA values ($\ell_{\rm ds}\simeq150$~bps and $p= 10.4$~bps). The ssDNA persistence length is $\ell_{\rm ss} = 3.7$~nm, compatible with experimental measurement~\cite{Tinland-Macro1997}, even though in the upper range of measured values.

The evolution of $\textbf{r}_i(t)$ is governed by the overdamped Langevin equation, 
integrated using a Euler's scheme,
\begin{equation}
\zeta \frac{d\textbf{r}_i}{dt} = -\nabla_{\textbf{r}_i}\mathcal{H}({\textbf{r}_j}) + \mathbf{\xi}(t) ,
\end{equation}
where $\zeta=3\pi\eta a$ is the friction coefficient for each bead of diameter $a$ with 
$\eta=10^{-3}$ Pa.s the water viscosity. 
The diffusion coefficient, $D_\textrm{diff} \equiv k_BT_0/3\pi\eta a$, thus takes into account 
the level of coarse-graining of the mesoscopic model involved in the kinetics associated 
to the smoothed free-energy landscape~\cite{Murtola-PCCP2009}. 
The random force of zero mean $\mathbf{\xi}_i(t)$ obeys 
the fluctuation-dissipation relation $\langle \mathbf{\xi}_i(t).\mathbf{\xi}_i(t')\rangle =6k_BT\zeta\delta_{ij}\delta(t-t')$. 
Lengths and energies are made dimensionless in the units of $a=0.34$ nm and $k_BT_0$, respectively. 
The dimensionless time step is $\delta\tau = \delta t k_B T_0/(a^2\zeta)$, set to $5 \times 10^{-4}$ 
($\delta t=0.045$ ps) for sufficient accuracy~\cite{Dasanna-EPL2012, Dasanna-PRE2013}. This set of parameters induces zipping velocities $v \approx 0.2-2$ bp/ns, compatible with experimental 
measurements~\cite{Bustamante-COSB2000}.
\begin{figure*}
\includegraphics[width=0.9\textwidth, angle=-0]{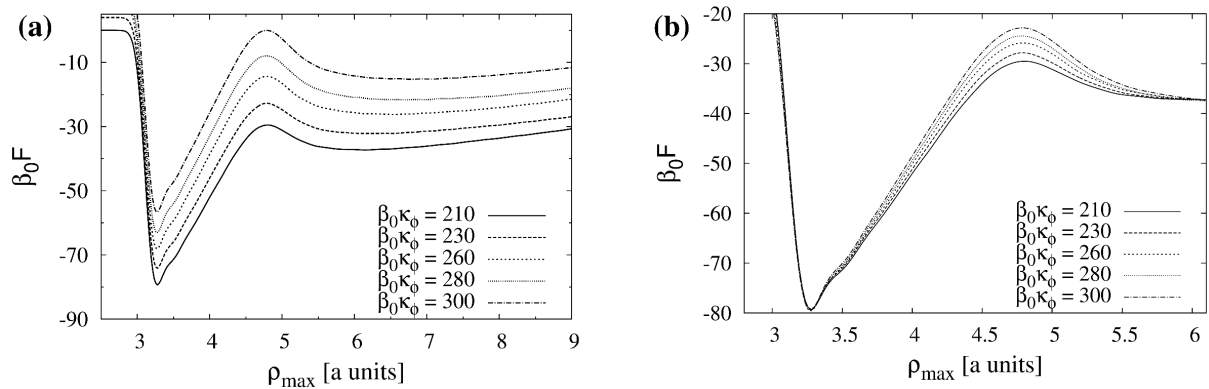}
 \caption{Evolution of the free-energy profile associated with the temperature-activated closure mechanism 
 for increasing values of $\beta_0\kappa_\phi$, and $\rho_b=1.50$ nm. (a) The free-energy 
 profiles are shifted arbitrarily along the ordinate axis for clarity. (b) The 
 free-energy profiles are fitted with respect to the closed-state basin to underline the increase of the barrier 
 height with $\beta_0\kappa_\phi$.}
 \label{figA3}
\end{figure*}
\begin{figure*}
\includegraphics[width=0.9\textwidth, angle=-0]{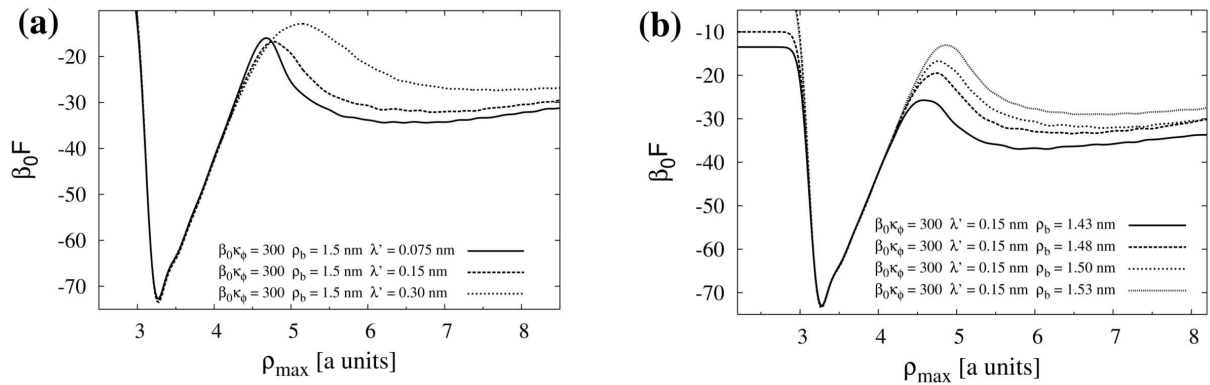}
 \caption{Evolution of the free-energy profile associated with the temperature-activated closure mechanism for $\beta_0 \kappa_\phi =300$ 
 reconstructed with Well-Tempered MetaDynamics, and for different values of the parameters $\lambda'$ and $\rho_b$ 
 (cf. Eq.~(\ref{stacking})). 
 We see that a slight change in these values does not change qualitatively the physics of the model.
 In the left panel (a), the closure free-energy barrier $\beta_0\Delta F_{\rm cl} = 18.6$, $15.4$, and $14.3$ 
 for $\lambda'= 0.075$ nm, $0.15$ nm, and $0.30$ nm. 
 In the right panel (b), $\beta_0\Delta F_{\rm cl} = 11.2$, $13.9$, $14.7$ and $15.6$ 
 for $\rho_b = 1.43$ nm, $1.48$ nm, $1.50$ nm and $1.53$ nm, respectively.}
 \label{figA4}
\end{figure*}

\section{Melting Temperature}
\begin{figure*}
 \includegraphics[width=0.75\textwidth, angle=-0]{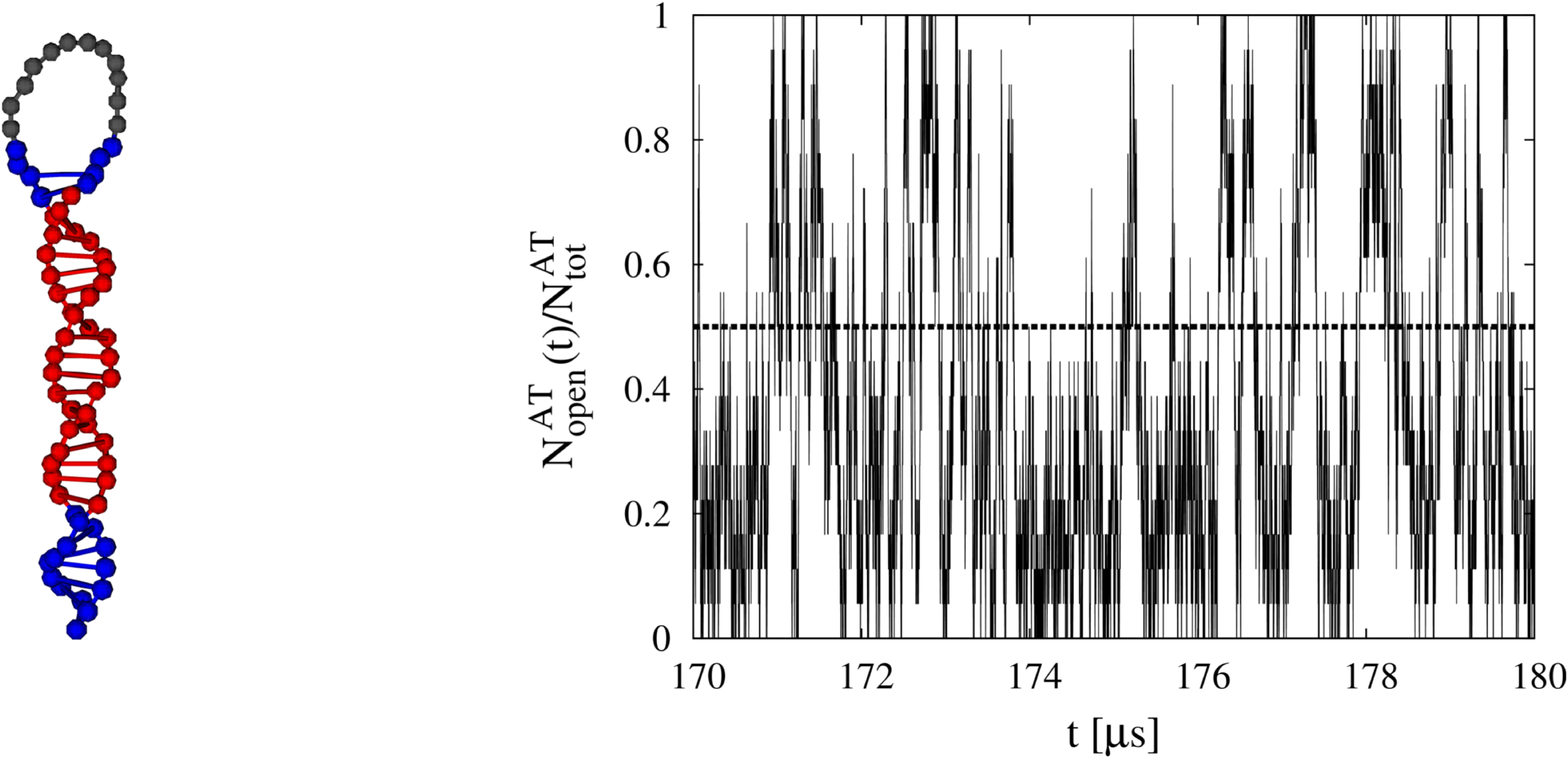}
 \caption{Left panel: Snapshot of an equilibrated DNA beacon of sequence $5'-GCGCG(AT)_9GCGC(T)_{12}CGCG(AT)_9CGCGC-3'$, 
 analogous to the sequence considered in Ref.~\onlinecite{Altan-PRL2003}, 
 in a closed configuration. The GC-rich regions are colored in blue, the middle AT-base-pair region in red, 
 and the loop T-rich region in gray. Right panel: illustration of the temporal evolution of the number 
 of opened AT-bps, $N_{\rm open}^{\rm AT}(t)$, at $T=T_{\rm m}^{\rm AT}$, associated with the internal-tagging. 
 The dashed line represents half of the number  of AT-bps in the  middle domain, 
 $N_{\rm open}^{\rm AT}(t)/N_{\rm tot}^{\rm AT} = 1/2$. We consider the middle AT-domain and the terminal GC-domain 
 to measure the melting temperatures $T_{\rm m}^{\rm AT}$ and $T_{\rm m}^{\rm GC}$, respectively.}
 \label{figA5}
\end{figure*}
Following Refs.~\onlinecite{Bonnet-PNAS1998, Altan-PRL2003}, we focus on the melting dynamics considering the mesoscopic model 
and DNA beacon configurations (see left panel in Fig.~\ref{figA5}). We consider a system made of $N=35$ beads 
with a different base sequence from the one considered in the main text to study bubble denaturation:
a GC-rich region at the extremity, a AT-base-pair region in the middle, and a T-rich region 
for the loop. It is indeed analogous to the sequence $5'-GCGCG(AT)_9GCGC(T)_{12}CGCG(AT)_9CGCGC-3'$ considered in Ref.~\onlinecite{Altan-PRL2003},  
even if the mesoscopic model does not distinguish explicitly the bases A and T (G and C respectively). 
The length of the loop region has been chosen such that it is greater than the single-stand persistence length 
$\ell_{\rm ss} \approx 10$ bps.

We study the melting properties associated with end-tagging (GC clamp) and internal-tagging (AT domain), 
\textit{i.e.} the number of open/closed bps as a function of the temperature. 
We thus assume that the DNA melting occurs when the average number of bps in the denaturation domain under consideration 
(AT or GC) is comparable with half of the number of DNA bps in this domain (see right panel in Fig.~\ref{figA5}). 
Considering the parameters $\beta_0\kappa_\phi = 580$ and $\rho_b = 1.20$ nm, that mainly control the 
width of the equilibrium well and the height of the energy barrier, one obtains $T_{\rm m}^{\rm AT} \approx 330$~K
 and $T_{\rm m}^{\rm GC} \approx 338$~K. 
It is well known that the melting temperature is sensitive to the salt concentration in the system.
Then, if we consider that the \textit{implicit} salt condition of our system is representative of a screened system, 
our result is compatible with experimental measurement, $T_{\rm m}^{\rm internal tag} \approx 345$~K
 and  $T_{\rm m}^{\rm end tag} \approx 350$~K. \cite{Altan-PRL2003}

\section{MetaDynamics Simulations}
\textbf{Thermodynamic properties}. Because of its convergence properties, Well-Tempered MetaDynamics (WT-metaD) 
is the most widely adopted version of the metadynamcis algorithm \cite{Barducci-PRL2008}. In WT-metaD, the bias deposition 
rate decreases over simulation time and the dynamics of all the microscopic variables becomes 
progressively closer to thermodynamic equilibrium as the simulation proceeds, making the bias 
to converge to its limiting value in a single run and avoiding the problem of overfilling, \textit{ie.} 
when the height of the accumulated Gaussians largely exceeds the true barrier height. 
Thus one prevents the system from being irreversibly pushed in 
regions of configuration space which are not physically relevant. 
Its success depends on the critical choice of a reasonable number of relevant collective variables (CVs). 
All the relevant slow varying degrees of freedom must be described by the CVs. In addition, the number of CVs 
must be small enough to avoid exceedingly long computational time, while being able to distinguish 
among the different conformational states of the system. 
However, to correctly describe the free-energy landscape, it is not necessary that the CVs chosen in metaD 
properly account for all the states and barriers. Actually, they must mainly account for the 
\textit{relevant} barriers associated with coarse-grained variables on which the free-energy dependence 
is the most important.
In the particular case of the slow closure mechanism of bubble denaturation studied in this Letter with 
the mesoscopic model, several observables come out to describe the metastable state as well 
as the transition to the closed state: (1) the length $L(t)$ of the bubble, 
\textit{i.e.} the number of opened base-pairs, (2) the width $\rho_{\max}(t)$ of the bubble, 
\textit{i.e.} the maximal distance between paired bases, (3) the average twist angle per bp, $\Delta \phi$, 
in the bubble \cite{Dasanna-PRE2013}, and (4) the minimal twist angle, $\phi_{\min}(t)$, 
in the bubble. As shown in Fig.~\ref{figA6} and Fig.~\ref{figA7}, the observables $\rho_{\max}(t)$ and 
$\phi_{\min}(t)$ are highly correlated to the length of the bubble, $L(t)$. 
This property is due to the relative simplicity of the mesoscopic model for which the stacking interaction 
is internally described.
\begin{figure*}
 \includegraphics[width=0.9\textwidth, angle=-0]{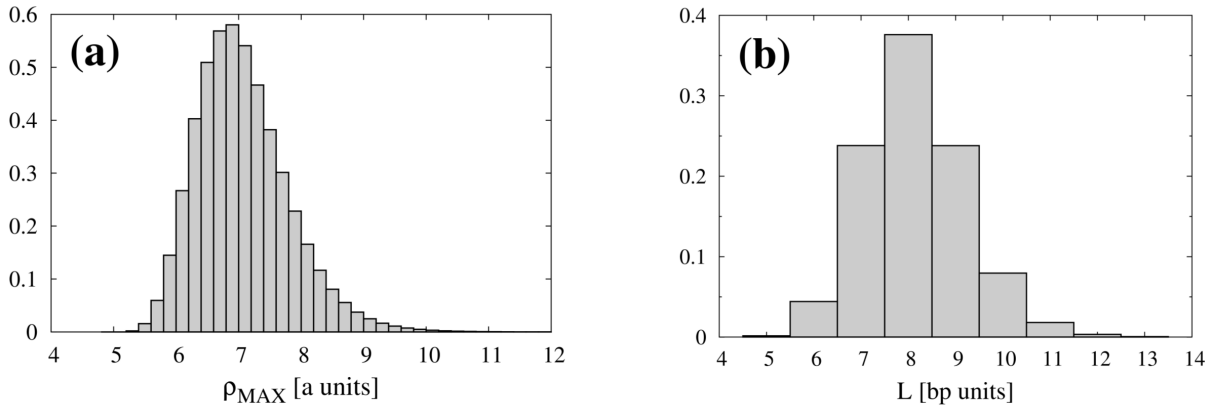}
 \caption{Distribution of the width $\rho_{\max}$ (a) and the length $L$ (b) of the metastable 
 bubble, respectively, for $\beta_0\kappa_\phi = 300$ and $\rho_b = 1.50$ nm, performed over a $60\, \mu$s unbiased trajectory 
 at room temperature.}
 \label{figA6}
\end{figure*}
\begin{figure*}
 \includegraphics[width=0.9\textwidth, angle=-0]{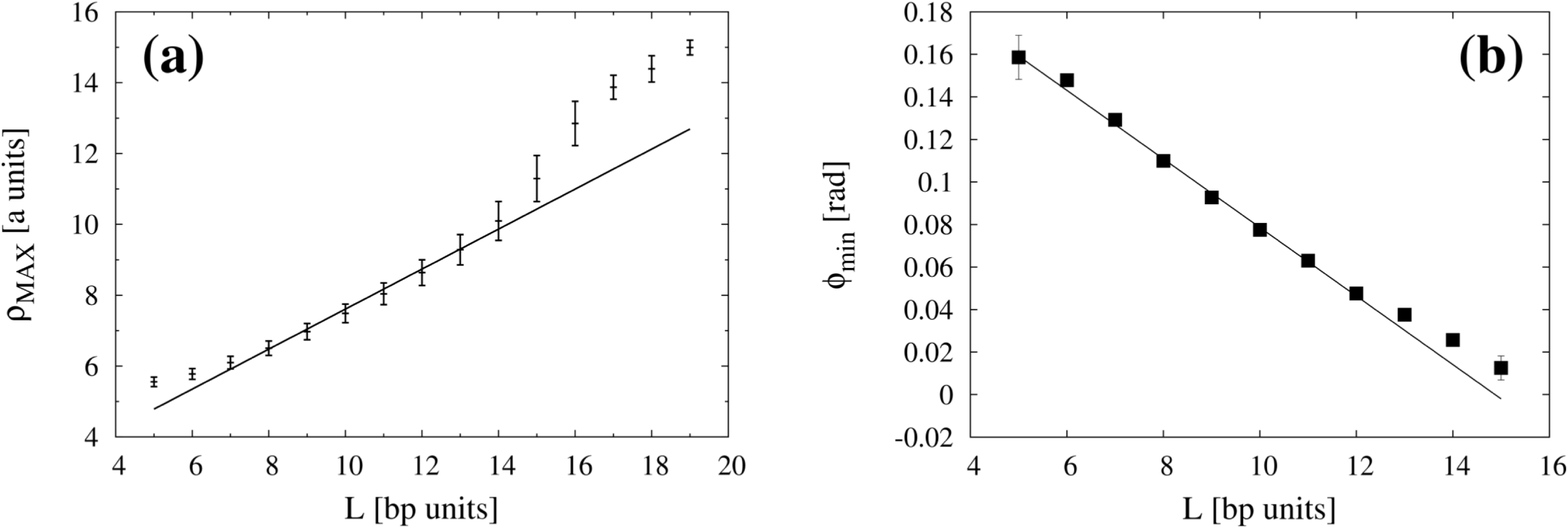}
 \caption{Average width $\rho_{\max}$ (a) and average minimal twist angle $\phi_{\min}$ (b) 
 of the metastable bubble as a function of the number of opened base-pairs $L$, for $\beta_0\kappa_\phi = 300$ and $\rho_b = 1.50$ nm,
 and performed over a $60\, \mu$s unbiased trajectory at room temperature (with standard deviation). 
 We emphasize the linear interpolation in the range of $L$ (respectively $\rho_{\max}$ and $\phi_{\min}$) 
 defining the metastable bubble.}
 \label{figA7}
\end{figure*}
Thus, a natural candidate for a relevant CV could be either the length $L(t)$ of the bubble 
itself, either its width $\rho_{\max}(t)$. In the following we choose for numerical efficiency 
the width $\rho_{\max}(t)$ as CV to bias the dynamics of the system.
According to the algorithm introduced by Barducci \textit{et al.}~\cite{Barducci-PRL2008,Bonomi-JCC2009} a Gaussian is deposited every $\tau_G = 25$ ps 
with height $w = w_0 e^{-V(s,t)/(f-1)T}$, where $s$ is the CV, $w_0 = 0.1\,k_B T$ is the initial height, $T$ is the temperature 
of the simulation, $V(s,t)$ the metadynamics time-dependent bias
\be
V(s,t)=\omega \sum_{t'<t} \exp\left[-\frac{(s(t)-s(t'))^2}{2\sigma^2}\right]
\ee
and $f\equiv (T+\Delta T)/T=5$ is the bias factor 
with $\Delta T$ a parameter with the dimension of a temperature. 
The resolution of the recovered free-energy landscape is determined by the width 
of the Gaussians $\sigma = 0.1$ in units of the respective CV.
Considering preliminary results of unbiased simulations (see Fig.~\ref{figA6}), we put a wall 
at $\rho_{\max} \approx 10$ to prevent the system to escape from the metastable state~\cite{Bonomi-CPC2009}
(and therefore entering in the zipping regime, \textit{i.e.} a far from equilibrium process~\cite{Dasanna-EPL2012, Dasanna-PRE2013}). 
We have checked that a slight change in the position of the wall ($\rho_{\max}=9,10,11,16,20$) does not change significantly 
the results, particularly the positions of the local minimum and the saddle, as well as the barrier height.
The simulations are run until the free-energy profile does not change more than 2 $k_B T$ 
in the last 100 ns. 
To further control the error of the reconstructed landscape 
we performed 5 runs of WT-metaD for each values of the parameter $\kappa_\phi$. 
The other observables are reconstructed afterwards using the \textit{reweighting technique} 
of Bonomi et al. \cite{Bonomi-JCC2009}.
Biased simulations were performed using the version 1.3 of the plugin for free-energy calculation, named PLUMED \cite{Bonomi-CPC2009}.\\

\textbf{Dynamical properties}. In order to estimate the mean transition times between the metastable (bubble) 
and the equilibrium (closed) states, we extend the \textit{standard} application scope  of metaD considering 
the recent method of Parrinello, Salvalaglio and Tiwary~\cite{Tiwary-PRL2013, Salvalaglio-JCTC2014}. 
We denote by $\tau$ the mean transition time over the barrier from the metastable state to the closed state, and 
by $\tau_M$ the mean transition time for the metadynamics run. The later changes as the simulation progresses and is 
linked to the former through the acceleration factor $\alpha(t) \equiv \langle e^{\beta V(s,t)} \rangle_M = \tau/\tau_M(t)$, 
where the angular brackets $\langle \dots \rangle_M$ denote an average over a metadynamics run confined to the metastable basin, 
and $V(s,t)$ is the metadynamics time-dependent bias. To satisfy the main validity criterions, \textit{ie.} 
1) to consider a set of CVs able to distinguish between the different metastable states~\cite{Salvalaglio-JCTC2014}, 
and 2) to avoid depositing bias in the Transition State region~\cite{Tiwary-PRL2013}, we check that 
the statistics of transition times follows 
a Poisson distribution (performing a two-sample Kolmogorov-Smirnov test with $p$-values in a range $[0.59,0.96]$), 
and increase the time lag between two successive Gaussian depositions $\tau_G = 600$ ps. 
We performed several WT-metaD simulations and stop the simulations when the crossing of the barrier and the Gaussian 
deposition occur unlikely at the same time. 
We have checked that the position of the wall does not affect the mean escape time, when put at a distance 
greater than the one defining the upper value of the metastable state, \textit{i.e.} $\rho_{\rm wall} > 10$ (see Fig.~\ref{figA6}). 
Actually, increasing significantly the time lag between 2 successive Gaussian depositions, one checks that the system 
is weakly perturbed at the border of the metastable basin only.
This analysis thus highlights the value of $\beta\kappa_\phi = 540$ associated with characteristic time scales, 
$\tau_{\rm cl} \approx 40~\rm{\mu s}$ and $\tau_{\rm nuc} \approx 15~\rm{ms}$, for the closure and nucleation 
mechanisms, in good agreement with experiments~\cite{Altan-PRL2003}.\\

\textbf{Measure of the effective torsional modulus, $\kappa_{\phi}^*(L)$}.
Considering the equipartition theorem, we measure 
$\kappa_{\phi}^* = k_{\rm B} T/ \langle (\Phi - \langle\Phi\rangle)^2 \rangle$, 
where $\Phi \equiv \sum_{i\in \textrm{bubble}} \phi_i$ is the twist angle 
measured \textit{consecutively} between the bps defining each extremity 
of the bubble and $L$ is the length of the bubble. We clearly see in Fig.~\ref{figA8} that $\kappa_{\phi}^*$ 
increases when $L$ decreases, recovering the value of the torsional modulus 
in the double-stranded domain, $\kappa^*_{\phi,\textrm{ds}}\simeq470~k_{\rm B}T_0$ 
(for $\beta_0\kappa_\phi=580$). Figure~\ref{figA8} highlights a non-trivial power law behaviour, 
$\kappa_{\phi}^*(L) \propto L^{-\alpha}$ with $\alpha = 2.2 \pm 0.1$. This law is valid 
down to $L \approx 3$~bps that corresponds to the breathing bubble regime. 
The origin of the free-energy barrier is indeed related to the finite value 
of $\kappa_{\phi}^*(L)$ in the metastable bubble and the crossover between 
two minima  for the minimal twist angle, $\phi^{\rm eq}_{\min}$ and 0.\\
\begin{figure}[t]
\includegraphics[width=0.9 \columnwidth]{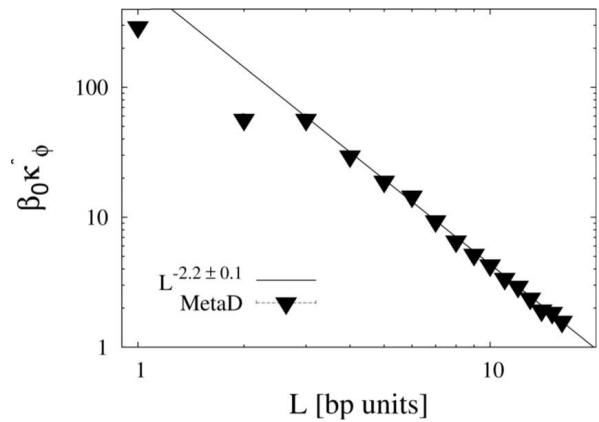}
 \caption{Bubble torsional modulus $\kappa_{\phi}^*(L)$ of the metastable bubble 
 computed using a standard metadynamics simulation with a wall at $\rho_{\max} \approx 20$.}
 \label{figA8}
\end{figure}

\textbf{Computation of the minimal free-energy path}. To obtain a typical minimal 
free-energy path (as in Fig.~1 in the main text), we applied the following methodology. We considered the free-energy 
surface reconstructed with Well-Tempered Metadynamics. We then fixed an \textit{initial point} 
located around the saddle region, \textit{i.e.} $\rho_{\max}=1.35$ nm and $\phi_{\min}=0.35$. Finally, we explored 
the trajectories associated with this initial point, and that drive the system to the metastable and equilibrium 
basins. Due to the inherent discretization scheme, this path is defined with relative accuracy (bin-size effect).


\begin{thebibliography}{99}
%
\bibitem{Kornberg-Freeman1992} A. Kornberg and T.A. Baker, \textit{DNA Replication} (W.H. Freeman, New York, 1992).
\bibitem{Phillips-Garland2009} R. Phillips, J. Kondev, J. Theriot, \textit{Physical Biology of the Cell} (Garland Science, 2009).
\bibitem{Leger-PNAS1998} J.F. L\'eger, J. Robert, L. Bourdieu, D. Chatenay, and J.F. Marko, Proc. Natl. Acad. Sci. U.S.A. \textbf{95}, 12295 (1998).
\bibitem{Kowalski-PNAS1988} D. Kowalski, D.A. Natale, and M.J. Eddy, Proc. Natl. Acad. Sci. U.S.A. \textbf{85}, 9464 (1988).
\bibitem{SantaLucia-PNAS1998} J. SantaLucia, Proc. Natl. Acad. Sci. U.S.A. \textbf{95}, 1460 (1998).
\bibitem{Poland-Academic1970} D. Poland and H.R. Scheraga, \textit{Theory of Helix Coil Transition in Biopolymers} (Academic Press, New York, 1970).
\bibitem{Benham1979} C.J. Benham, Proc. Natl. Acad. Sci. U.S.A. \textbf{76}, 3870 (1979).
\bibitem{Jeon-PRL2010} J-H. Jeon, J. Adamcik, G. Dietler, and R. Metzler, Phys. Rev. Lett. \textbf{105}, 208101 (2010).
\bibitem{Manghi-JPCM2009} M. Manghi, J. Palmeri, and N. Destainville, J. Phys.: Condens. Matter \textbf{21}, 034104 (2009).
\bibitem{Dasanna-PRE2013} A.K. Dasanna, N. Destainville, J. Palmeri, and M. Manghi, Phys. Rev. E \textbf{87}, 052703 (2013).
\bibitem{Altan-PRL2003} G. Altan-Bonnet, A. Libchaber, and O. Krichevsky, Phys. Rev. Lett. \textbf{90}, 138101 (2003).
\bibitem{Cheatham-JMB1996} T.E. Cheatham and P.A. Kollman, J. Mol. Biol. \textbf{259}, 434 (1996).
\bibitem{Dixit-BJ2005} S.B. Dixit, D.L. Beveridge, D.A. Case, T.E. Cheatham, E. Giudice, F. Lankas, R. Lavery, J.H. Maddocks, R. Osman, H. Sklenar, K.M. Thayer, and P. Varnai, Biophys. J. \textbf{89}, 3721 (2005).
\bibitem{Kannan-PCCP2009} S. Kannan and M. Zacharias, Phys. Chem. Chem. Phys. \textbf{11}, 10589 (2009).
\bibitem{Sayar-PRE2010} M. Sayar, B. Av\c{s}aro\u{g}lu, and A. Kabak\c{c}io\u{g}lu, Phys. Rev. E \textbf{81}, 041916 (2010).
\bibitem{Ouldridge-JCP2011} T.E. Ouldridge, A.A. Louis, and J.P. Doye, J. Chem. Phys. \textbf{134}, 085101 (2011).
\bibitem{Savelyev-PNAS2010} A. Savelyev and G.A. Papoian, Proc. Natl. Acad. Sci. U.S.A. \textbf{107}, 20340 (2010).
\bibitem{Englander-PNAS1980} S.W. Englander, N.R. Kallenbach, A.J. Heeger, J.A. Krumhansl, and S. Litwin, Proc. Natl. Acad. Sci. U.S.A. \textbf{77}, 7222 (1980).
\bibitem{Zeida-PRE2012} A. Zeida, M.R. Machado, P.D. Dans, and S. Pantano, Phys. Rev. E \textbf{86}, 021903 (2012).
\bibitem{Peyrard-PRL1989} M. Peyrard and A.R. Bishop, Phys. Rev. Lett. \textbf{62}, 2755 (1989).
\bibitem{Barbi-PL1999} M. Barbi, S. Cocco, and M. Peyrard, Phys. Lett. A \textbf{253}, 5178 (1999).
\bibitem{Jeon-JCP2006} J.-H. Jeon, W. Sung, and F.H. Ree, J. Chem. Phys. \textbf{124}, 164905 (2006).
\bibitem{Bar-PRL2007} A. Bar, Y. Kafri, and D. Mukamel, Phys. Rev. Lett. \textbf{98}, 038103 (2007).
\bibitem{Fogedby-PRL2007} H.C. Fogedby and R. Metzler, Phys. Rev. Lett. \textbf{98}, 070601 (2007).
\bibitem{Mielke-JCP2005} S. P. Mielke, N. Gronbech-Jensen, V. V. Krishnan, W. H. Fink, and C. J. Benham, J. Chem. Phys. \textbf{123}, 124911 (2005).
\bibitem{Laio-PNAS2002} A. Laio and M. Parrinello, Proc. Natl. Acad. Sci. U.S.A. \textbf{99}, 12562 (2002).
\bibitem{Laio-RPP2008} A. Laio and F.L. Gervasio, Rep. Prog. Phys. \textbf{71}, 126601 (2008).
\bibitem{Barducci-PRL2008} A. Barducci, G. Bussi, and M. Parrinello, Phys. Rev. Lett. \textbf{100}, 020603 (2008).
\bibitem{Dama-PRL2014} J. F. Dama, M. Parrinello, and G. A. Voth, Phys. Rev. Lett. \textbf{112}, 240602 (2014).
\bibitem{Anil-Thesis2003} A.K. Dasanna, Ph.D. thesis, Toulouse University (2013).
\bibitem{Warmlander-Biochem1999} S. W\"{a}rml\"{a}nder, A. Sen, and M. Leijon, Biochemistry \textbf{39}, 607 (1999).
\bibitem{Bauer-JMB1993} W. R. Bauer and C. J. Benham, J. Mol. Biol. \textbf{234}, 1184 (1993).
\bibitem{Vlaminck-MC2012} I. De Vlaminck, M. T. J. van Loenhout, L. Zweifel, J. den Blanken, K. Hooning, S. Hage, J. Kerssemakers, and C. Dekker, Molecular Cell \textbf{46}, 616 (2012).
\bibitem{Tiwary-PRL2013} P. Tiwary and M. Parrinello, Phys. Rev. Lett. \textbf{111}, 230602 (2013).
\bibitem{Salvalaglio-JCTC2014} M. Salvalaglio, P. Tiwary, and M. Parrinello, J. Chem. Theory Comput. \textbf{10}, 1420 (2014).
\bibitem{Qi-eLife2013} Z. Qi, R.A. Pugh, M. Spies, and Y.R. Chemla, eLife \textbf{2}, e00334 (2013).
\bibitem{Myong-Science2007} S. Myong, M.M. Bruno, A.M. Pyle, and T. Ha, Science \textbf{317}, 513 (2007).
\bibitem{Grosberg-AIP1994} A.Y. Grosberg and A.R. Khokhlov, \textit{Statistical Physics of Macromolecules} (AIP, Melville, NY, 1994).
\bibitem{footnote} The length of the AT-rich region only plays a role during the zipping process, and does not influence the height of the activation barriers, as long as it is greater than the typical size of the metastable bubble (see Ref.~\onlinecite{Dasanna-PRE2013}). Considering smaller AT tracts shifts the departure from the metastable basin to lower $\rho_{\textrm{max}}$, consequently decreasing the height of the closure free-energy barrier.
\bibitem{Bryant-Nature2003} Z. Bryant, M.D. Stone, J. Gore, S.B. Smith, N.R. Cozzarelli, and C. Bustamante, Nature \textbf{142}, 338 (2003).
\bibitem{Dasanna-EPL2012} A.K. Dasanna, N. Destainville, J. Palmeri, and M. Manghi, EPL \textbf{98}, 38002 (2012).
\bibitem{Bonomi-JCC2009} M. Bonomi, A. Barducci, and M. Parrinello, J. Comput. Chem. \textbf{30}, 1615 (2009).
\bibitem{Hanggi-RMP1990} P. H{\"a}nggi, P. Talkner, and M. Borkovec, Rev. Med. Phys. \textbf{62}, 251 (1990).
\bibitem{Bonnet-PNAS1998} G. Bonnet, O. Krichevsky, and A. Libchaber, Proc. Natl. Acad. Sci. U.S.A \textbf{95}, 8602 (1998).
\bibitem{Sponer-Biopolymers2002} J. Sponer, J. Leszczynski, and P. Hobza, Biopolymers \textbf{61}, 2 (2002).
\bibitem{Manghi-SM2006} M. Manghi, Y-W. Kim, X. Schlagberger, and R.R. Netz, Soft Matter \textbf{2}, 653 (2006).
\bibitem{Liu-PNAS1987} L. F. Liu and J. C. Wang, Proc. Natl. Acad. Sci. U.S.A. \textbf{84}, 7024 (1987).
\bibitem{Michelotti-MCB1996} G.A. Michelotti, E.F. Michelotti, A. Pullner, R.C. Duncan, D. Eick, and D. Levens, Mol. Cell. Biol. \textbf{16}, 2656 (1996).
\bibitem{Kouzine-Cell2013} F. Kouzine, D. Wojtowicz, A. Yamane, W. Resch, K.-R. Kieffer-Kwon, R. Bandle, S. Nelson, H. Nakahashi,  P. Awasthi, L. Feigenbaum, H. Menoni, J. Hoeijmakers, W. Vermeulen, H. Ge, T. M. Przytycka, D. Levens, and R. Casellas, Cell \textbf{153}, 988 (2013).
\bibitem{Bikard-MMBR2010} D. Bikard, C. Loot, Z. Baharoglu, and D. Mazel, Microbiol. Mol. Biol. Rev. \textbf{74}, 570 (2010).
\bibitem{Alexandrov-PLOS2009} B.S. Alexandrov, V. Gelev, S.W. Yoo, A.R. Bishop, K. Rasmussen, and A. Usheva, PLoS Comput. Biol. \textbf{5}, e1000313 (2009).
\bibitem{Jeon-BJ2008} J.-H. Jeon, W. Sung, Biophys. J. \textbf{95}, 3600 (2008).
\bibitem{Bonomi-CPC2009} M. Bonomi, D. Branduardi, G. Bussi, C. Camilloni, D. Provasi, P. Raiteri, D. Donadjo, F. Marinelli, F.Pietrucci, R.A. Broglia, and M. Parrinello, Comput. Phys. Comm. \textbf{180}, 1961 (2009).
\bibitem{Hugel-PRL2005} T. Hugel, M. Rief, M. Seitz, H.E. Gaub, and R. Netz, Phys. Rev. Lett. \textbf{94}, 048301 (2005).
\bibitem{Tinland-Macro1997} B. Tinland, A. Pluen, J. Sturm, and G. Weill, Macromol. \textbf{30}, 5763 (1997).
\bibitem{Murtola-PCCP2009} T. Murtola, A. Bunker, I. Vattulainen, M. Deserno, and M. Karttunen, Phys. Chem. Chem. Phys. \textbf{11}, 1869 (2009).
\bibitem{Bustamante-COSB2000} C. Bustamante, S.B. Smith, J. Liphardt, and D. Smith, Curr. Opin. Struct. Biol. \textbf{10}, 279 (2000).

\end{thebibliography}
\end{document}